\errorstopmode
\input amssym.def
\input amssym.tex


\magnification=\magstephalf
\hsize=14.0 true cm
\vsize=19 true cm
\hoffset=1.0 true cm
\voffset=2.0 true cm

\abovedisplayskip=12pt plus 3pt minus 3pt
\belowdisplayskip=12pt plus 3pt minus 3pt
\parindent=1.0em


\font\sixrm=cmr6
\font\eightrm=cmr8
\font\ninerm=cmr9

\font\sixi=cmmi6
\font\eighti=cmmi8
\font\ninei=cmmi9

\font\sixsy=cmsy6
\font\eightsy=cmsy8
\font\ninesy=cmsy9

\font\sixbf=cmbx6
\font\eightbf=cmbx8
\font\ninebf=cmbx9

\font\eightit=cmti8
\font\nineit=cmti9

\font\eightsl=cmsl8
\font\ninesl=cmsl9

\font\sixss=cmss8 at 8 true pt
\font\sevenss=cmss9 at 9 true pt
\font\eightss=cmss8
\font\niness=cmss9
\font\tenss=cmss10

\font\sixmib=cmmib6
\font\sevenmib=cmmib7
\font\eightmib=cmmib8
\font\ninemib=cmmib9
\font\tenmib=cmmib10

 at 12 true pt
 at 12 true pt
\font\bigrm=cmr10 at 12 true pt
 at 12 true pt
 at 12 true pt

 at 16 true pt
 at 16 true pt
\font\Bigrm=cmr12 at 16 true pt
 at 16 true pt
 at 16 true pt

\catcode`@=11
\newfam\ssfam
\newfam\mibfam

\def\tenpoint{\def\rm{\fam0\tenrm}%
    \textfont0=\tenrm \scriptfont0=\sevenrm \scriptscriptfont0=\fiverm
    \textfont1=\teni  \scriptfont1=\seveni  \scriptscriptfont1=\fivei
    \textfont2=\tensy \scriptfont2=\sevensy \scriptscriptfont2=\fivesy
    \textfont3=\tenex \scriptfont3=\tenex   \scriptscriptfont3=\tenex
    \textfont\itfam=\tenit                  \def\it{\fam\itfam\tenit}%
    \textfont\slfam=\tensl                  \def\sl{\fam\slfam\tensl}%
    \textfont\bffam=\tenbf \scriptfont\bffam=\sevenbf
                           \scriptscriptfont\bffam=\fivebf
                           \def\bf{\fam\bffam\tenbf}%
    \textfont\ssfam=\tenss \scriptfont\ssfam=\sevenss
                           \scriptscriptfont\ssfam=\sevenss
                           \def\ss{\fam\ssfam\tenss}%
    \textfont\mibfam=\tenmib \scriptfont\mibfam=\sevenmib
                             \scriptscriptfont\mibfam=\sevenmib
                             \def\mib{\fam\mibfam\tenmib}%
    \normalbaselineskip=13pt
    \setbox\strutbox=\hbox{\vrule height8.5pt depth3.5pt width0pt}%
    \let\big=\tenbig
    \normalbaselines\rm}

\def\ninepoint{\def\rm{\fam0\ninerm}%
    \textfont0=\ninerm      \scriptfont0=\sixrm
                            \scriptscriptfont0=\fiverm
    \textfont1=\ninei       \scriptfont1=\sixi
                            \scriptscriptfont1=\fivei
    \textfont2=\ninesy      \scriptfont2=\sixsy
                            \scriptscriptfont2=\fivesy
    \textfont3=\tenex       \scriptfont3=\tenex
                            \scriptscriptfont3=\tenex
    \textfont\itfam=\nineit \def\it{\fam\itfam\nineit}%
    \textfont\slfam=\ninesl \def\sl{\fam\slfam\ninesl}%
    \textfont\bffam=\ninebf \scriptfont\bffam=\sixbf
                            \scriptscriptfont\bffam=\fivebf
                            \def\bf{\fam\bffam\ninebf}%
    \textfont\ssfam=\niness \scriptfont\ssfam=\sixss
                            \scriptscriptfont\ssfam=\sixss
                            \def\ss{\fam\ssfam\niness}%
    \textfont\mibfam=\ninemib \scriptfont\mibfam=\sixmib
                            \scriptscriptfont\mibfam=\sixmib
                            \def\mib{\fam\mibfam\ninemib}%
    \normalbaselineskip=12pt
    \setbox\strutbox=\hbox{\vrule height8.0pt depth3.0pt width0pt}%
    \let\big=\ninebig
    \normalbaselines\rm}

\def\eightpoint{\def\rm{\fam0\eightrm}%
    \textfont0=\eightrm      \scriptfont0=\sixrm
                             \scriptscriptfont0=\fiverm
    \textfont1=\eighti       \scriptfont1=\sixi
                             \scriptscriptfont1=\fivei
    \textfont2=\eightsy      \scriptfont2=\sixsy
                             \scriptscriptfont2=\fivesy
    \textfont3=\tenex        \scriptfont3=\tenex
                             \scriptscriptfont3=\tenex
    \textfont\itfam=\eightit \def\it{\fam\itfam\eightit}%
    \textfont\slfam=\eightsl \def\sl{\fam\slfam\eightsl}%
    \textfont\bffam=\eightbf \scriptfont\bffam=\sixbf
                             \scriptscriptfont\bffam=\fivebf
                             \def\bf{\fam\bffam\eightbf}%
    \textfont\ssfam=\eightss \scriptfont\ssfam=\sixss
                             \scriptscriptfont\ssfam=\sixss
                             \def\ss{\fam\ssfam\eightss}%
    \textfont\mibfam=\eightmib \scriptfont\mibfam=\sixmib
                             \scriptscriptfont\mibfam=\sixmib
                             \def\mib{\fam\mibfam\eightmib}%
    \normalbaselineskip=10pt
    \setbox\strutbox=\hbox{\vrule height7.0pt depth2.0pt width0pt}%
    \let\big=\eightbig
    \normalbaselines\rm}

\def\tenbig#1{{\hbox{$\left#1\vbox to8.5pt{}\right.\n@space$}}}
\def\ninebig#1{{\hbox{$\textfont0=\tenrm\textfont2=\tensy
                       \left#1\vbox to7.25pt{}\right.\n@space$}}}
\def\eightbig#1{{\hbox{$\textfont0=\ninerm\textfont2=\ninesy
                       \left#1\vbox to6.5pt{}\right.\n@space$}}}

\font\sectionfont=cmbx10
\font\subsectionfont=cmti10

\def\figurecaptionfont{\ninepoint}
\def\tablecaptionfont{\ninepoint}
\def\footnotefont{\eightpoint}


\newcount\equationno
\newcount\bibitemno
\newcount\figureno
\newcount\tableno

\equationno=0
\bibitemno=0
\figureno=0
\tableno=0


\footline={\ifnum\pageno=0{\hfil}\else
{\hss\rm\the\pageno\hss}\fi}


\def\section #1. #2 \par
{\vskip0pt plus .10\vsize\penalty-100 \vskip0pt plus-.10\vsize
\vskip 1.6 true cm plus 0.2 true cm minus 0.2 true cm
\global\def\equationlabel{#1}
\global\equationno=0
\leftline{\sectionfont #1. #2}\par
\immediate\write\terminal{Section #1. #2}
\vskip 0.7 true cm plus 0.1 true cm minus 0.1 true cm
\noindent}


\def\subsection #1 \par
{\vskip0pt plus 0.8 true cm\penalty-50 \vskip0pt plus-0.8 true cm
\vskip2.5ex plus 0.1ex minus 0.1ex
\leftline{\subsectionfont #1}\par
\immediate\write\terminal{Subsection #1}
\vskip1.0ex plus 0.1ex minus 0.1ex
\noindent}


\def\appendix #1. #2 \par
{\vskip0pt plus .10\vsize\penalty-100 \vskip0pt plus-.10\vsize
\vskip 1.6 true cm plus 0.2 true cm minus 0.2 true cm
\global\def\equationlabel{\hbox{\rm#1}}
\global\equationno=0
\leftline{\sectionfont Appendix #1. #2}\par
\immediate\write\terminal{Appendix #1. #2}
\vskip 0.7 true cm plus 0.1 true cm minus 0.1 true cm
\noindent}



\def\equation#1{$$\displaylines{\qquad #1}$$}
\def\enum{\global\advance\equationno by 1
\hfill\llap{{\rm(\equationlabel.\the\equationno)}}}
\def\noenum{\hfill}

\def\nexteq#1{\cr\noalign{\vskip#1}\qquad}


\def\ifundefined#1{\expandafter\ifx\csname#1\endcsname\relax}

\def\ref#1{\ifundefined{#1}?\immediate\write\terminal{unknown reference
on page \the\pageno}\else\csname#1\endcsname\fi}

\newwrite\terminal
\newwrite\bibitemlist

\def\bibitem#1#2\par{\global\advance\bibitemno by 1
\immediate\write\bibitemlist{\string\def
\expandafter\string\csname#1\endcsname
{\the\bibitemno}}
\item{[\the\bibitemno]}#2\par}

\def\beginbibliography{
\vskip0pt plus .15\vsize\penalty-100 \vskip0pt plus-.15\vsize
\vskip 1.2 true cm plus 0.2 true cm minus 0.2 true cm
\leftline{\sectionfont References}\par
\immediate\write\terminal{References}
\immediate\openout\bibitemlist=biblist
\frenchspacing\parindent=1.8em
\vskip 0.5 true cm plus 0.1 true cm minus 0.1 true cm}

\def\endbibliography{
\immediate\closeout\bibitemlist
\nonfrenchspacing\parindent=1.0em}


\def\figurecaption#1{\global\advance\figureno by 1
\narrower\figurecaptionfont
Fig.~\the\figureno. #1}

\def\tablecaption#1{\global\advance\tableno by 1
\vbox to 0.25 true cm { }
\centerline{\tablecaptionfont%
Table~\the\tableno. #1}
\vskip-0.4 true cm}

\def\thicktablerule{\hrule height0.8pt}
\def\thintablerule{\hrule height0.4pt}

\tenpoint

\immediate\openin\bibitemlist=biblist
\ifeof\bibitemlist\immediate\closein\bibitemlist
\else\immediate\closein\bibitemlist
\input biblist \fi


\def\thismonth{\ifcase\month\or
January\or February\or March\or April\or May\or June\or
July\or August\or September\or October\or November\or December\fi}

\input epsf
\epsfclipon



\def\rmd{{\rm d}}
\def\rmD{{\rm D}}
\def\rme{{\rm e}}
\def\rmO{{\rm O}}


\def\Re{{\rm Re}\,}


\def\proof{\noindent{\sl Proof:}\kern0.6em}

\def\frac#1#2{\hbox{$#1\over#2$}}
\def\dual{\mathstrut^*\kern-0.1em}

\def\lvec#1{\setbox0=\hbox{$#1$}
    \setbox1=\hbox{$\scriptstyle\leftarrow$}
    #1\kern-\wd0\smash{
    \raise\ht0\hbox{$\raise1pt\hbox{$\scriptstyle\leftarrow$}$}}
    \kern-\wd1\kern\wd0}
\def\rvec#1{\setbox0=\hbox{$#1$}
    \setbox1=\hbox{$\scriptstyle\rightarrow$}
    #1\kern-\wd0\smash{
    \raise\ht0\hbox{$\raise1pt\hbox{$\scriptstyle\rightarrow$}$}}
    \kern-\wd1\kern\wd0}
\def\slash#1{\setbox0=\hbox{$#1$}\setbox1=\hbox{$\kern1pt/$}
    #1\kern-\wd0\kern1pt/\kern-\wd1\kern\wd0}


\def\nabstar#1{{\nabla\kern0.5pt\smash{\raise 4.5pt\hbox{$\ast$}}
               \kern-5.5pt_{#1}}}

\def\drvstar#1{{\partial\kern0.5pt\smash{\raise 4.5pt\hbox{$\ast$}}
               \kern-6.0pt_{#1}}}

\def\ldrvstar#1{{\lvec{\,\partial}\kern-0.5pt\smash{\raise 4.5pt\hbox{$\ast$}}
               \kern-5.0pt_{#1}}}


\def\MeV{{\rm MeV}}

\def\MSbar{\overline{\rm MS\kern-0.5pt}\kern0.5pt}


\def\euler{\gamma_{\rm E}}



\def\diracstar#1#2{
    \setbox0=\hbox{$\gamma$}\setbox1=\hbox{$\gamma_{#1}$}
    \gamma_{#1}\kern-\wd1\kern\wd0
    \smash{\raise4.5pt\hbox{$\scriptstyle#2$}}}


\def\SUthree{{\rm SU(3)}}
\def\SUn{{\rm SU}(N)}

\def\suthree{\frak{su}(3)}
\def\sun{\frak{su}(N)}
\def\tr{{\rm tr}}


\def\eps{\epsilon}
\def\Sw{S_{\rm w}}
\def\Nf{N_{\rm f}}

\def\Obs{{\cal O}}

%
\rightline{CERN-PH-TH/2010-143}

\vskip1.5cm 
\centerline{\Bigrm
Properties and uses of the Wilson flow in lattice QCD
}
%
\vskip 0.6 true cm
\centerline{\bigrm Martin L\"uscher}
\vskip1.5ex
\centerline{{\it CERN, Physics Department, 1211 Geneva 23, Switzerland}}
\vskip 0.8 true cm
\thintablerule
\vskip 2.0ex
\ninepoint
\leftline{\bf Abstract}
\vskip 1.0ex\noindent
Theoretical and numerical studies of the Wilson flow 
in lattice QCD suggest that the gauge field obtained
at flow time $t>0$ is a smooth renormalized field.
The expectation values of local gauge-invariant
expressions in this field are thus well-defined physical quantities
that probe the theory at length scales on the order of $\sqrt{t}$.
Moreover, by transforming the QCD functional
integral to an integral over the gauge field at a specified
flow time, the emergence of the topological (instanton)
sectors in the continuum limit becomes transparent and
is seen to be caused by a dynamical effect that
rapidly separates the sectors
when the lattice spacing is reduced from $0.1$ fm
to smaller values.
\vskip 2.0ex
\thintablerule

\tenpoint

\vskip-0.3cm

\section 1. Introduction

Flows in field space are an interesting tool that may
allow new insights to be gained into the physical 
mechanisms described
by highly non-linear quantum field theories such as QCD. 
The flow $B_{\mu}(t,x)$ of $\SUthree$ gauge fields 
studied in this paper is 
defined by the equations
\equation{
  \dot{B}_{\mu}=D_{\nu}G_{\nu\mu},
  \qquad
  \left.B_{\mu}\right|_{t=0}=A_{\mu},
  \enum
  \nexteq{2.5ex}
  G_{\mu\nu}=\partial_{\mu}B_{\nu}-\partial_{\nu}B_{\mu}
  +[B_{\mu},B_{\nu}],
  \qquad
  D_{\mu}=\partial_{\mu}+[B_{\mu},\,\cdot\;],
  \enum
}
where $A_{\mu}$ is the fundamental gauge field in QCD
(see Appendix A for unexplained notation; differentiation with 
respect to the ``flow time'' $t$ is abbreviated by a dot). 
Evidently, for increasing $t$ and as long as no singularities
develop, the flow equation (1.1) drives the gauge field
along the direction of steepest descent towards the 
stationary points of the Yang--Mills action. 

In lattice QCD, the simplest choice of the action
of the gauge field $U(x,\mu)$ is the Wilson action
[\ref{Wilson}]
\equation{
  \Sw(U)={1\over g_0^2}\sum_p
  \Re\tr\{1-U(p)\},
  \enum
}
where $g_0$ is the bare coupling,
$p$ runs over all oriented plaquettes on the lattice 
and $U(p)$ denotes the product of 
the link variables around $p$.
The associated flow $V_t(x,\mu)$ of 
lattice gauge fields (the ``Wilson flow'')
is defined by the equations
\equation{
  \dot{V}_t(x,\mu)=-g_0^2\left\{\partial_{x,\mu}\Sw(V_t)\right\}V_t(x,\mu),
  \qquad
  \left.V_t(x,\mu)\right|_{t=0}=U(x,\mu),
  \enum
}
in which $\partial_{x,\mu}$ stands for the natural $\suthree$-valued
differential operator with respect to the 
link variable $V_t(x,\mu)$ (see Appendix A).
The existence, uniqueness and smoothness of the Wilson flow at 
all positive and negative times $t$ is rigorously guaranteed on 
a finite lattice [\ref{ArnoldI}]. Moreover, from eq.~(1.4)
one immediately concludes that
the action $\Sw(V_t)$ is a monotonically decreasing function of $t$.
The flow therefore tends to have a smoothing
effect on the field and it is, in fact, generated by infinitesimal
``stout link smearing'' steps [\ref{Stout}].

The Wilson flow previously appeared in ref.~[\ref{TrivMaps}] 
in the context
of trivializing maps of field space.
Familiarity with this paper is not assumed, but some 
mathematical results obtained there will be used here again.
An important goal in the following is to find out whether
the expectation values of local observables constructed 
from the gauge field at positive flow time can be expected 
to have a well-defined continuum limit.
Evidence for the existence of the limit is provided
by performing a sample calculation to one-loop order 
of perturbation theory directly in the continuum theory, using
dimensional regularization, and through a
numerical study of the $\SUthree$ gauge theory at three values
of the lattice spacing.
Two applications of the Wilson flow are then discussed, 
one concerning the scale-setting in lattice QCD and the
other the question of
how exactly the topological (instanton) sectors emerge 
when the lattice spacing goes to zero.

\section 2. Properties of the Wilson flow at small coupling

The aim in this section partly is to show that the 
Wilson flow can be studied straightforwardly in perturbation
theory and partly to check that
the expectation values of local gauge-invariant 
observables calculated at positive flow time 
are renormalized quantities.

For simplicity the perturbation expansion
is discussed in the continuum theory using
dimensional regularization.
The gauge group is taken to be $\SUn$
and it is assumed that there are $\Nf$ flavours of 
massless quarks. As a representative case,
the observable
\equation{
   E=\frac{1}{4}G_{\mu\nu}^aG_{\mu\nu}^a
   \enum
}
is considered and its expectation value is worked out to 
next-to-leading order in the gauge coupling.

\subsection 2.1 Gauge fixing

The flow equation (1.1) is invariant 
under $t$-independent gauge transformations.
$E$ is therefore a gauge-invariant function of the 
fundamental field $A_{\mu}$ and its expectation value
can consequently be calculated in any gauge.

In perturbation theory, the gauge invariance of 
the flow equation leads to some
technical complications that are better avoided by 
considering the modified equation
\equation{
  \dot{B}_{\mu}=D_{\nu}G_{\nu\mu}+\lambda D_{\mu}\partial_{\nu}B_{\nu}.
  \enum
}
For any given value of the gauge parameter $\lambda$,
the solution of eq.~(2.2) is related to the one at 
$\lambda=0$ through
\equation{
  B_{\mu}=\Lambda\!\left.B_{\mu}\right|_{\lambda=0}\Lambda^{-1}
  +\Lambda\partial_{\mu}\Lambda^{-1},
  \enum
}
where the gauge transformation $\Lambda(t,x)$ is determined by
\equation{
  \dot{\Lambda}=-\lambda \partial_{\nu}B_{\nu}\Lambda,
  \qquad\left.\Lambda\right|_{t=0}=1.
  \enum
}
The expectation value of $E$ 
can thus be computed using the modified 
flow equation. Moreover, one is free to set $\lambda=1$, which
turns out to be a particularly convenient choice.

Note that the use of the modified flow equation does not interfere
with the fixing of the gauge of the fundamental field, since
$E$ is unchanged and therefore remains a gauge-invariant
function of the latter.

\subsection 2.2 Solution of the modified flow equation

In perturbation theory the gauge potential is scaled by the
bare coupling,
\equation{
  A_{\mu}\to g_0A_{\mu},
  \enum
}
and the functional integral is then expanded in powers of $g_0$.
The flow $B_{\mu}(t,x)$ thus becomes a function
of the coupling with an asymptotic expansion of the form
\equation{
  B_{\mu}=\sum_{k=1}^{\infty}g_0^kB_{\mu,k},
  \qquad
  \left.B_{\mu,k}\right|_{t=0}=\delta_{k1}A_{\mu}.
  \enum
}
When this series is inserted in eq.~(2.2),
and if one sets $\lambda=1$,
a tower of equations
\equation{
  \dot{B}_{\mu,k}-\partial_{\nu}\partial_{\nu}B_{\mu,k}=
  R_{\mu,k},
  \quad k=1,2,\ldots,
  \enum
}
is obtained, where the expressions on the right are given by
\equation{
  R_{\mu,1}=0,
  \enum
  \nexteq{2ex}
  R_{\mu,2}=2[B_{\nu,1},\partial_{\nu}B_{\mu,1}]
            -[B_{\nu,1},\partial_{\mu}B_{\nu,1}],
  \enum
  \nexteq{2ex}
  R_{\mu,3}=2[B_{\nu,2},\partial_{\nu}B_{\mu,1}]
            +2[B_{\nu,1},\partial_{\nu}B_{\mu,2}]
  \noenum
  \nexteq{1.5ex}
  {\phantom{R_{\mu,3}=}}
            -[B_{\nu,2},\partial_{\mu}B_{\nu,1}]
            -[B_{\nu,1},\partial_{\mu}B_{\nu,2}]
            +[B_{\nu,1},[B_{\nu,1},B_{\mu,1}]],
  \enum
}
and so on. In particular, in $D$ dimensions
the leading-order equation implies
\equation{
  B_{\mu,1}(t,x)=\int\rmd^Dy\,K_t(x-y)A_{\mu}(y),
  \enum
  \nexteq{2.5ex}
  K_t(z)=\int{\rmd^Dp\over(2\pi)^D}\,\rme^{ipz}\rme^{-tp^2}=
  {\rme^{-z^2/4t}\over(4\pi t)^{D/2}},
  \enum
}
which shows explicitly that the flow is a smoothing operation.
More precisely, the gauge potential is averaged over a 
spherical range in space whose 
mean-square radius in four dimensions is
equal to $\sqrt{8t}$.

The higher-order equations (2.7) can be solved one after another 
by noting that
\equation{
   B_{\mu,k}(t,x)=\int_0^t\rmd s\int\rmd^Dy\,K_{t-s}(x-y)R_{\mu,k}(s,y).
   \enum
}
Recalling eqs.~(2.9),(2.10), it is clear that this
formula generates tree-like expressions, where the fundamental field
$A_{\mu}$ is attached to the endpoints of the trees.

\subsection 2.3 Expansion of $\langle E\rangle$

When the series (2.6) is inserted in 
\equation{
  \langle E\rangle=
           \frac{1}{2}\langle\partial_{\mu}B^a_{\nu}\partial_{\mu}B^a_{\nu}-
           \partial_{\mu}B^a_{\nu}\partial_{\nu}B^a_{\mu}\rangle
  \noenum
  \nexteq{2.5ex}
  {\phantom{\langle E\rangle=}}
           +f^{abc}\langle\partial_{\mu}B^a_{\nu}B^b_{\mu}B^c_{\nu}\rangle
           +\frac{1}{4}f^{abe}f^{cde}\langle
           B^a_{\mu}B^b_{\nu}B^c_{\mu}B^d_{\nu}\rangle,
  \enum
}
a sequence of terms of increasing order in $g_0$ is obtained.
The lowest-order term is
\equation{
  {\cal E}_0=\frac{1}{2}g_0^2
             \langle\partial_{\mu}B^a_{\nu,1}\partial_{\mu}B^a_{\nu,1}-
             \partial_{\mu}B^a_{\nu,1}\partial_{\nu}B^a_{\mu,1}\rangle
  \enum
}
and the terms at the next order are
\equation{
  {\cal E}_1=g_0^3
    f^{abc}\langle\partial_{\mu}B^a_{\nu,1}B^b_{\mu,1}B^c_{\nu,1}\rangle,
  \enum
  \nexteq{2.5ex}
  {\cal E}_2=
    g_0^3
    \langle\partial_{\mu}B^a_{\nu,2}\partial_{\mu}B^a_{\nu,1}-
           \partial_{\mu}B^a_{\nu,2}\partial_{\nu}B^a_{\mu,1}\rangle.
  \enum
}
Each of these terms is a power series in the gauge coupling,
which may be worked out by expressing the 
coefficients $B_{\mu,k}(t,x)$ through the fundamental 
field $A_{\mu}(x)$ and by expanding 
the correlation functions of the latter using
the standard Feynman rules.

In practice it is advantageous to pass to momentum space by
inserting the Fourier representations
\equation{
  B^a_{\mu,1}(t,x)=\int_p\rme^{ipx}\rme^{-tp^2}\tilde{A}^a_{\mu}(p),
  \enum
  \nexteq{2.5ex}
  B^a_{\mu,2}(t,x)=if^{abc}\int_0^t\rmd s\int_{q,r}
  \rme^{i(q+r)x}\rme^{-s(q^2+r^2)-(t-s)(q+r)^2}
  \noenum
  \nexteq{2ex}
  {\phantom{B^a_{\mu,2}(t,x)=}}
  \times
  \bigl\{\delta_{\mu\lambda}r_{\sigma}-\delta_{\mu\sigma}q_{\lambda}
  +\frac{1}{2}\delta_{\sigma\lambda}(q-r)_{\mu}\bigr\}
  \tilde{A}_{\sigma}^b(q)\tilde{A}_{\lambda}^c(r),
  \enum
}
and the corresponding
expressions for the higher-order fields $B_{\mu,k}(t,x)$.
The shorthand
\equation{
  \int_{p}=\int{\rmd^D p\over(2\pi)^D}
  \enum
}
has been introduced here in order to simplify the notation. 

Note that the flow-time integral in eq.~(2.19) 
is similar to a Feynman parameter integral.
In particular, 
if Feynman parameters are used for
the diagrams contributing to the gluon
correlation functions,
one ends up with gaussian momentum integrals 
that can be easily evaluated in any dimension $D$.
The integrals over the parameters then look very much
the same as the ones usually encountered,
except for the fact that the flow-time parameters
are integrated up to $t$ rather than infinity.

\subsection 2.4 Computation of ${\cal E}_0$

In the case of the lowest-order term (2.15), 
the steps sketched in the previous subsection 
lead to the formula
\equation{
  {\cal E}_0=\frac{1}{2}g_0^2(N^2-1)
  \int_p\rme^{-2tp^2}
  (p^2\delta_{\mu\nu}-p_{\mu}p_{\nu})
  D(p)_{\mu\nu},
  \enum
}
where $D(p)_{\mu\nu}$ denotes the unrenormalized full gluon propagator.
Setting $D=4-2\eps$ and choosing the Feynman gauge,
the propagator assumes the form
\equation{
  D(p)_{\mu\nu}={1\over(p^2)^2}\left\{
  (p^2\delta_{\mu\nu}-p_{\mu}p_{\nu})(1-\omega(p))^{-1}+p_{\mu}p_{\nu}
  \right\},
  \enum
  \nexteq{2.5ex}
  \omega(p)=\sum_{k=1}^{\infty}g_0^{2k}(p^2)^{-k\eps}\omega_k,
  \enum
}
from which one infers that
\equation{
  {\cal E}_0=\frac{1}{2}g_0^2{N^2-1\over(8\pi t)^{D/2}}(D-1)
  \left\{1+g_0^2(2t)^{\eps}{\Gamma(2-2\eps)\over\Gamma(2-\eps)}\omega_1
  +\ldots\right\}.
  \enum
}
To this order, the computation is then easily completed by
quoting the known result
\equation{
  \omega_1={1\over 16\pi^2}(4\pi\rme^{-\euler})^{\eps}
  \left\{N\left(\frac{5}{3\eps}+\frac{31}{9}\right)
  -\Nf\left(\frac{2}{3\eps}+\frac{10}{9}\right)
  +\rmO(\eps)\right\}
  \enum
}
for the gluon self-energy, $\euler=0.577\ldots$ being
Euler's constant.

\subsection 2.5 Computation of $\langle E\rangle$ to order $g_0^4$

At the next-to-leading order, the expectation value of $E$
receives contributions
from ${\cal E}_0$ and the terms 
${\cal E}_1,{\cal E}_2,\ldots$ 
up to order $g_0^4$
generated by the expansion of 
the flow equation
(cf.~subsect.~2.3). Some of the latter involve 
the $3$-point gluon vertex, but in all cases only the
leading-order gluon correlation functions are required.

In the case of the term ${\cal E}_2$, for example, 
the purely algebraic part of the calculation leads to
the integral
\equation{
   {\cal E}_2=N(N^2-1)g_0^4\int_0^t\rmd s\int_{q,r}
   {\rme^{-(2t-s)p^2-s(q^2+r^2)}\over p^2q^2r^2}
   \noenum
   \nexteq{1.5ex}
   \quad\times\bigl\{
   (D-1)p^2(p^2+q^2+r^2)+2(D-2)(q^2r^2-(qr)^2)\bigr\}
   +\rmO(g_0^6)
   \enum
}
where $p=q+r$. This integral appears to be quite complicated, but
the polynomial in the numerator of the integrand allows the 
expression to be simplified and eventually to be evaluated
analytically (see Appendix B for further details).
The result of the computation,
\equation{
   {\cal E}_2=N(N^2-1){g_0^4\over(4\pi)^D(2t)^{D-2}}
   \bigl\{\frac{9}{2\eps}-\frac{3}{2}+45\ln2-\frac{45}{2}\ln3+
   \rmO(\eps)\bigr\}+\rmO(g_0^6),
   \noenum
   \nexteq{0.0ex}
   \enum
}
shows that these contributions are not free of ultra-violet 
singularities.

The computation of the other terms
follows the same pattern and does not
present any additional difficulties.
Collecting all contributions, the result
\equation{
  \langle E\rangle=\frac{1}{2}g_0^2{N^2-1\over(8\pi t)^{D/2}}(D-1)
  \bigl\{1+c_1g_0^2+\rmO(g_0^4)
  \bigr\},
  \enum
  \nexteq{2.5ex}
  c_1={1\over16\pi^2}(4\pi)^{\eps}(8t)^{\eps}\bigl\{
  N\left(\frac{11}{3\eps}+\frac{52}{9}-3\ln3\right)
  -\Nf\left(\frac{2}{3\eps}+\frac{4}{9}-\frac{4}{3}\ln2\right)
  +\rmO(\eps)\bigr\},
  \noenum
  \nexteq{0.0ex}
  \enum
}
is then obtained.

\subsection 2.6 Renormalization

The bare coupling $g_0$ is related to 
the renormalized coupling $g$ 
in the $\MSbar$ scheme  
and the associated normalization mass $\mu$
according to [\ref{BardeenEtAl}]
\equation{
  g_0^2=g^2\mu^{2\eps}(4\pi\rme^{-\euler})^{-\eps}
  \Bigl\{1-\frac{1}{\eps}b_0g^2+\rmO(g^4)\Bigr\},
  \enum
  \nexteq{2ex}
  b_0={1\over16\pi^2}\left\{\frac{11}{3}N-\frac{2}{3}\Nf\right\}.
  \enum
}
Now if eq.~(2.28) is written as an expansion in the renormalized coupling,
the terms proportional to $1/\eps$ cancel and one obtains
\equation{
  \langle E\rangle={3(N^2-1)g^2\over128\pi^2t^2}
  \bigl\{1+\bar{c}_1g^2+\rmO(g^4)
  \bigr\},
  \enum
  \nexteq{2.5ex}
  \bar{c}_1=
  {1\over16\pi^2}\bigl\{
  N\left(\frac{11}{3}L+\frac{52}{9}-3\ln3\right)
  -\Nf\left(\frac{2}{3}L+\frac{4}{9}-\frac{4}{3}\ln2\right)\bigr\},
  \enum
}
at $\eps=0$, where $L=\ln(8\mu^2t)+\euler$.
To this order in the gauge coupling, 
$\langle E\rangle$ thus turns out to be a
quantity that does not need to be renormalized
and which therefore encodes some physical property of 
the theory.

In terms of the running coupling $\alpha(q)$
at scale $q=(8t)^{-1/2}$,
the expansion (2.32) assumes the form
\equation{
  \langle E\rangle={3(N^2-1)\over32\pi t^2}\alpha(q)
  \bigl\{1+k_1\alpha(q)+\rmO(\alpha^2)
  \bigr\},
  \enum
  \nexteq{2.5ex}
  k_1=
  {1\over4\pi}\bigl\{
  N\left(\frac{11}{3}\euler+\frac{52}{9}-3\ln3\right)
  -\Nf\left(\frac{2}{3}\euler+\frac{4}{9}-\frac{4}{3}\ln2\right)\bigr\}.
  \enum
}
In particular, for $N=3$ one obtains
\equation{
  \langle E\rangle={3\over4\pi t^2}\alpha(q)
  \bigl\{1+k_1\alpha(q)+\rmO(\alpha^2)
  \bigr\},
  \qquad
  k_1=1.0978+0.0075\times\Nf.
  \enum
}
The next-to-leading order correction is reasonably small in 
this case and corresponds
to a change in the momentum $q$ by a factor $2$ or so if $\Nf\leq3$.

\section 3. Lattice studies of the Wilson flow

In QCD the perturbation expansion (2.36)
is expected to be applicable
at small flow times only, where the smoothing range
$\sqrt{8t}$ is at most $0.3$ fm or so.
The properties of the Wilson flow at larger values of $t$ can however
be studied straightforwardly using the lattice formulation of the
theory and numerical simulations.

The simulations of the $\SUthree$ gauge theory 
reported in this section mainly serve to clarify whether
$\langle E\rangle$ scales to the continuum limit as
suggested by perturbation theory. Along the way, 
a new scale-setting method will be proposed based on the 
observed properties of $\langle E\rangle$.

\topinsert
\newdimen\digitwidth
\setbox0=\hbox{\rm 0}
\digitwidth=\wd0
\catcode`@=\active
\def@{\kern\digitwidth}
\tablecaption{Lattice parameters, statistics and reference flow time}
\vskip-1.0ex
$$\vbox{\settabs\+&%
                  xxxxxxxxxxxx&xx&
                  xxxxxxx&xx&
                  xxxxxxxxxxxx&xx&
                  xxxxxxx&xx&
                  xxxxxxxxxxxxx&x\cr
\thicktablerule
\vskip1.0ex
                \+& \hfill Lattice\hfill
                 && \hfill $\beta$\hfill
                 && \hfill $a$\kern2pt[fm]\hfill
                 && \hfill $N_{\rm cnfg}$\hfill
                 && \hfill $t_0/a^2$\hfill
                 &\cr
\vskip1.0ex
\thintablerule
\vskip1.2ex
  \+& \hfill $48\times24^3$\hfill
  &&  \hfill $5.96$\hfill
  &&  \hfill $0.0999(4)$\hfill
  &&  \hfill $100$ \hfill
  &&  \hfill $@2.7854(62)$\hfill
  &\cr
\vskip0.3ex
  \+& \hfill $64\times32^3$\hfill
  &&  \hfill $6.17$\hfill
  &&  \hfill $0.0710(3)$\hfill
  &&  \hfill $100$ \hfill
  &&  \hfill $@5.489(14)@$\hfill
  &\cr
\vskip0.3ex
  \+& \hfill $96\times48^3$\hfill
  &&  \hfill $6.42$\hfill
  &&  \hfill $0.0498(3)$\hfill
  &&  \hfill $100$ \hfill
  &&  \hfill $11.241(23)@$\hfill
  &\cr
\vskip1.2ex
\thicktablerule
}
$$
\endinsert

\subsection 3.1 Simulation parameters

Ensembles of representative gauge fields were 
generated on three lattices, using the Wilson gauge action
and a combination of the well-known link-update algorithms.
The values of the lattice spacing quoted in table~1 
derive from the results in lattice units for the Sommer 
reference scale $r_0=0.5$ fm [\ref{SommerScale}] 
published by Guagnelli et al.~[\ref{GuagnelliEtAl}].
At the chosen couplings $\beta=6/g_0^2$,
the spacings of the three lattices thus
decrease from roughly $0.1$ to $0.05$ fm by factors of $1/\sqrt{2}$.
Moreover, the lattice sizes in physical units are approximately
constant.

In order to safely suppress any residual statistical correlations of
the $N_{\rm cnfg}$ generated fields, the separation in simulation time
of the fields was taken to be at least $10$ times
the integrated autocorrelation time of the topological
charge. The definition of the latter on the lattice
is ambiguous to some extent, but 
its autocorrelation time is largely independent
of the choices one makes and is known to increase very rapidly
when the lattice spacing is reduced 
[\ref{DelDebbioTauQ},\ref{StefanConference}].
In particular, already at $a=0.07$ fm 
all other usual quantities of interest tend to be far less 
correlated in simulation time.

\subsection 3.2 Observables

For any given gauge field configuration $U(x,\mu)$, the flow equation (1.4)
can be integrated numerically up to the desired flow time and
one may then construct gauge-invariant local observables from 
the gauge field at this time. In particular, 
\equation{
  E=2\sum_{p\in P_x}\Re\tr\{1-V_t(p)\}
  \enum
}
is a possible definition of the density $E$ on the lattice,
$P_x$ being the set of unoriented plaquettes
with lower-left corner $x$.

\topinsert
\vbox{
\vskip0.0cm
\centerline{\epsfxsize=5.0cm\epsfbox{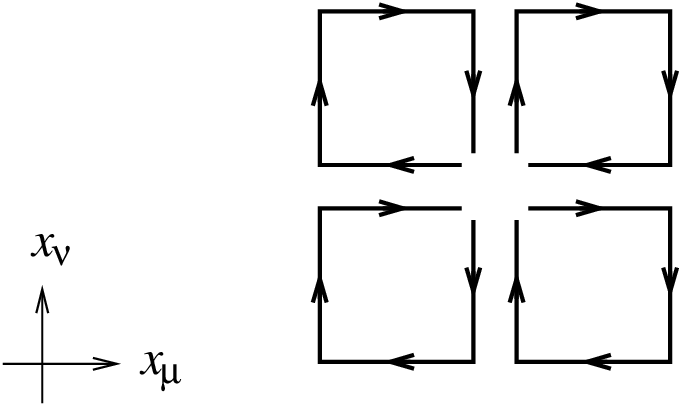}}
\vskip0.5cm
\figurecaption{%
The anti-hermitian traceless 
part of the average of the four plaquette Wilson loops
shown in this figure can be taken as the
definition of the field tensor $a^2G_{\mu\nu}(x)$
on the lattice. All loops are in the 
$(\mu,\nu)$-plane, have the same orientation
and start and end at the point $x$.
}
}
\endinsert

Another more symmetric definition of $E$ is 
obtained by introducing a lattice version of the 
field tensor $G_{\mu\nu}(x)$ (see fig.~1).
$E$ is then simply given by the continuum formula (2.1).
Both definitions 
are equally acceptable at this point,
since they respect all formal requirements 
(locality and gauge invariance in particular) and since
they converge to 
the correct expression in the classical continuum limit.

For the numerical integration of the flow equation (1.4), 
any of the widely known 
integration schemes can in principle be used 
(see ref.~[\ref{HairerEtAl}], for example).
The Euler integrators previously discussed in ref.~[\ref{TrivMaps}] 
are particularly simple but also the least efficient ones.
Evidently, the integration errors should be much smaller than
the statistical errors of the calculated quantities.
A fairly simple and numerically stable 
integrator that allows this condition to be easily met
is described in Appendix C.

\topinsert
\vbox{
\vskip0.0cm
\centerline{\epsfxsize=8.5cm\epsfbox{plots/tsqE.eps}}
\vskip0.5cm
\figurecaption{%
Simulation data obtained at $a=0.05$ fm for 
$t^2\langle E\rangle$ as a function of the flow time $t$
(black line). Statistical errors are smaller than $0.3\%$ 
and therefore invisible on the scale of the figure.
The curve predicted by the perturbation series (2.36) 
and the known value (3.2) of the $\Lambda$-parameter
is also shown (grey band).
}
}
\endinsert

\subsection 3.3 Time dependence of $\langle E\rangle$

To leading-order perturbation theory, the dimensionless combination 
$t^2\langle E\rangle$ is a constant proportional to the gauge coupling.
At the next order, the scale invariance of the theory 
is broken and $t^2\langle E\rangle$ develops a 
non-trivial dependence on 
the flow time $t$. Asymptotic freedom actually implies that
the combination slowly goes to zero in the limit $t\to0$.

Recalling the result
\equation
{
  \left.\Lambda\right|_{\Nf=0}=0.602(48)/r_0
  \enum
} 
for the $\Lambda$-parameter in the $\MSbar$ scheme
obtained by the {\ninepoint ALPHA} collaboration [\ref{LambdaParm}],
and using the four-loop evolution equation
for the running coupling $\alpha(q)$ [\ref{RitbergenEtAl}],
the perturbation series (2.36) can be 
evaluated at any value of the flow time
given in units of $r_0$. 
The curve obtained in this way
is shown in fig.~2 together with the error band that derives
from the error of $\Lambda$ quoted in eq.~(3.2). 

Perhaps somewhat fortuitously, the simulation results obtained
at $a=0.05$ fm accurately match the perturbative curve 
over a significant range of $t$. 
The symmetric definition of $E$ has here been used and 
for clarity the data from only one lattice are shown
(as discussed below, the lattice-spacing effects are small 
and the data from the other lattices would therefore lie nearly on 
top of the line plotted in fig.~2).

Beyond the perturbative regime, $t^2\langle E\rangle$ grows 
roughly linearly with $t$, at least so within the 
range covered by the simulation data. 
The slowdown of the density 
$\langle E\rangle$ from the
perturbative $1/t^2$ to a smoother $1/t$ behaviour
may perhaps be explained by noting that
the Wilson flow tends to drive the gauge field towards the 
stationary points of the gauge action. In the vicinity 
of these points of field space, the right-hand
side of the flow equation (1.4) is small and $E$
consequently changes only little with time.

\subsection 3.4 Lattice-spacing effects

Perturbation theory suggests that
the density $\langle E\rangle$ scales to the continuum 
limit like a physical quantity of dimension $4$.
The scaling behaviour of $\langle E\rangle$ can be
checked by introducing a reference scale $t_0$ through 
the implicit equation
\equation{
  \left\{t^2\langle E\rangle\right\}_{t=t_0}=0.3
  \enum
}
(see fig.~2). If $\langle E\rangle$ is physical,
the dimensionless ratio $t_0/r_0^2$ must be independent 
of the lattice spacing, up to corrections vanishing
proportionally to a power of $a$.

\topinsert
\vbox{
\vskip0.0cm
\centerline{\epsfxsize=8.5cm\epsfbox{plots/t0r0.eps}}
\vskip0.4cm
\figurecaption{%
Extrapolation of the dimensionless ratio $\sqrt{8t_0}/r_0$ of
reference scales to
the continuum limit (open data points). The black 
data points were obtained using the symmetric definition of
$E$ and the grey ones using the expression (3.1).
}
\vskip0.0cm
}
\endinsert

The data plotted in fig.~3 clearly show that the ratio of 
reference scales smoothly extrapolates to the continuum limit.
As may have been suspected, the lattice effects appear to 
be of order $a^2$. They are at most a few percent on the
lattices considered and particularly small
if the symmetric definition of $E$ is employed. 
At other points in time, the scaling violations behave similarly,
but tend to increase towards small $t$, where
the smoothing range $\sqrt{8t}$ is only 2 or 3 times larger than
the lattice spacing.

\subsection 3.5 Synthesis

(a) {\it Continuum limit}. 
The numerical studies 
of the Wilson flow strongly support the conjecture that
the gauge fields generated at positive flow time
are smooth renormalized fields (except for their gauge
degrees of freedom). In the case of QCD with a non-zero number
of sea quarks, similar studies however still need to be performed. 
Additional confirmation in perturbation theory and through
the consideration of further observables is evidently
desirable.

\vskip0.5ex
\noindent
(b) {\it Scale setting}. 
The time $t_0$ defined through eq.~(3.3) may serve as
a reference scale similar to the Sommer radius $r_0$.
With respect to the latter, $t_0$ has the advantage that
its computation does not require any fits or extrapolations.
Moreover, an ensemble of only 100 independent 
representative field configurations allows
$t_0$ to be obtained
with a statistical precision of a small fraction of a percent
(for illustration, the values of $t_0/a^2$ computed
using the symmetric definition of $E$ are listed in table~1).

\vskip0.5ex
\noindent
(c) {\it Universality}.
On the lattice the definition of
local gauge-invariant quantities like $E$ is not unique,
but the results obtained in this section indicate that
the differences become irrelevant in the continuum limit
(see fig.~3).
This kind of universality is a consequence of the fact,
further elucidated in sect.~4, that
the gauge fields generated by the Wilson flow are smooth
on the scale of the lattice spacing. 
The universality classes of the local fields composed from the 
gauge field at fixed $t/t_0$ 
are therefore determined by the asymptotic behaviour of the fields
in the classical continuum limit.

\section 4. Functional integral and topological sectors

In lattice gauge theory, the space of gauge fields is connected
and the concept of a topological sector has therefore no 
a priori well-defined meaning. 
However, one also knows since a long time that any classical
continuous gauge field (including multi-instanton configurations) can
be approximated arbitrarily well by lattice fields. 
The sectors are hence included in the field space
but are not separated
from one another.

An understanding of 
how exactly the sectors 
get divided when the lattice spacing is taken to zero
can now be achieved using the Wilson flow.
The existence of the topological sectors thus
turns out to be a dynamical property of the theory rather than
being a consequence of an assumed or imposed continuity 
of the fields.

\subsection 4.1 Field transformation

On a finite lattice, however large, the transformation
\equation{
  U\to V=V_{t_0}
  \enum
}
is invertible and actually a diffeomorphism of the field space
[\ref{TrivMaps}]. One can therefore perform a change of integration
variables in the QCD functional integral from the fundamental field
to the field at flow time $t_0$. 
As already noted in ref.~[\ref{TrivMaps}],
the asso\-ciated Jacobian can be worked out analytically and be
expressed through the Wilson action along the flow.
The expectation value of any observable $\Obs(V)$ is then
given by
\equation{
  \langle\Obs\rangle={1\over{\cal Z}}\int
  \rmD[V]\,\Obs(V)\,\rme^{-\tilde{S}(V)},
  \enum
  \nexteq{2.5ex}
  \tilde{S}(V)=S(U)+{16g_0^2\over3a^2}\int_0^{t_0}\rmd t\,\Sw(V_t),
  \qquad
  \rmD[V]=\prod_{x,\mu}\rmd V(x,\mu),
  \enum
}
where $S(U)$ denotes the total action 
(including the quark determinants) 
of the theory before the transformation.

Evidently, the functional integral may be rewritten in this way
as an integral over the gauge field at any flow time $t$.
Setting $t$ to the reference time $t_0$ is however an
interesting and natural choice. 
In particular, the fields that dominate
the transformed integral then have a characteristic
wavelength on the order of the fundamental 
low-energy scales of the theory.

\subsection 4.2 Smoothness of the dominant fields

A quantitative measure for the smoothness of a given lattice gauge 
field $V$ is
\equation{
  h=\max_{p}s_p,\qquad s_p=\Re\tr\{1-V(p)\},
  \enum
}
the maximum being taken over all plaquettes $p$
(as before, $V(p)$ denotes the product of the link variables 
around $p$).
In the functional integral (4.2), all possible values of $h$ occur,
but large values are expected to be unlikely
in view of the fact that the expectation value $\langle s_p\rangle$ 
scales proportionally to $a^4$.

\topinsert
\vbox{
\vskip0.0cm
\centerline{\epsfxsize=8.0cm\epsfbox{plots/prob.eps}}
\vskip0.4cm
\figurecaption{%
Plot of the probability for 
$s_p$ [eq.~(4.4)] on a given plaquette $p$ to be above
some specified value $s$.
From top to bottom, the three curves shown correspond to 
the values $a=0.1$, $0.07$ and $0.05$ fm 
of the lattice spacing.
The calculation was performed in the 
$\SUthree$ gauge theory using the ensembles
of fields described in subsect.~3.1.
}
\vskip0.0cm
}
\endinsert

Since all terms in the action (4.3)
have the same sign and favour smooth configurations,
it is certainly plausible that 
large plaquette values $s_p$ are strongly suppressed.
The probability for $s_p$ on a given plaquette $p$ to be larger than 
some specified value $s$ in fact decreases roughly like $a^{10}$
when the lattice spacing is reduced (see fig.~4).
In a finite volume of fixed physical size,
the statistical weight of the field configurations with 
values of $h$ beyond a 
given threshold is therefore rapidly going to zero
in the continuum limit, while the fields that dominate
the transformed functional integral
become uniformly smooth in space.

\subsection 4.3 Dynamical separation of the topological sectors

Many years ago, the space of
lattice gauge fields was shown to
divide into disconnected sectors
if only fields satisfying a certain smoothness
condition are admitted
[\ref{TopCharge},\ref{PhillipsStone}].
The proof is based on a geometrical construction 
of a local lattice expression 
for the topological charge which assumes integer values and 
which has the correct classical continuum limit.

In the case of QCD, the fields satisfying%
\kern1.5pt\footnote{$\dagger$}{\footnotefont%
The theorems proved by Phillips and Stone [\ref{PhillipsStone}] 
assume a simplicial lattice. To be able to apply them
in the present context, the hypercubic cells of the 
lattice must be divided 
into simplices. An interpolation of the gauge field to the
added links is then required and can easily be accomplished
using the same interpolation methods as the ones applied elsewhere
in refs.~[\ref{TopCharge},\ref{PhillipsStone}].
}
\equation{
   h<0.067
   \enum
}
are included in the subspace covered by the geometrical construction
and thus fall into topological sectors very much like
the continuous gauge fields in the continuum theory. The space of fields
``between the sectors'' may accordingly be characterized
by the inequality $h\geq 0.067$. 

Recalling the discussion in subsect.~4.2 
of the smoothness properties of the fields that dominate
the functional integral (4.2), the topological sectors
are now seen to emerge as a consequence of the fact that
the fields with large values of $h$ have a rapidly
decreasing weight in the integral when the lattice
spacing is taken zero. 
In the case of the simulations reported in sect.~3, for example,
the fraction of representative fields satisfying the bound (4.5)
increases from $0\%$ at $a=0.1$ fm to $8\%$ and $70\%$ at 
$a=0.07$ and $0.05$ fm, respectively. 
The asymptotic behaviour of the weight of the fields
between the sectors cannot be safely determined from these
data, but the curves shown in fig.~4 suggest that the 
probability of finding such a configuration 
on lattices of a given physical size 
goes to zero proportionally to $a^6$.

\subsection 4.4 Topological susceptibility

Once the functional integral (4.2) splits
into a sum of integrals over effectively 
disconnected sectors,
the assignment of the topological charge $Q$
to the field configurations 
in a representative ensemble of fields
becomes unambiguous.
The geometrical construction mentioned before
could be used to compute the charge, but
in view of the universality property
discussed at the end of sect.~3,
a straightforward discretization of 
the topological density, using the symmetric lattice 
expression for the field tensor $G_{\mu\nu}$,
is expected to give the same results for the moments
$\langle Q^n\rangle$ in the continuum limit. 

On the lattices listed in table~1,
the second moment $\langle Q^2\rangle$ 
computed along these lines is about $50$
and the topological susceptibility $\chi_t$ turns out
to be independent of the lattice spacing
within statistical errors. A fit by a constant
then gives the result
\equation{
   \chi_t^{1/4}=187.4(3.9)\,\MeV,
   \enum
}
which differs 
by less than two standard deviations
from the value $194.5(2.4)$ MeV [\ref{ChiGW}] 
obtained at flow time $t=0$
using a chiral lattice Dirac operator and
the index theorem [\ref{IndThm}].
Moreover, as shown in a forthcoming publication, 
the ratios of the fourth root of the second moments computed here
and through a variant [\ref{ChiProj}] of 
the universal formula proposed in 
[\ref{ChiInd}] extrapolate to $1$ in the continuum limit
to within a statistical uncertainty of $3\%$.

All these empirical results support the conjecture
that the moments of 
the charge distribution in the functional integral (4.2)
coincide with those obtained from the index theorem 
[\ref{IndThm}--\ref{ChiProj}] and thus the
ones appearing in 
the chiral Ward identities [\ref{ChiWI},\ref{ChiWII}].
However, while highly plausible, the equality remains to be 
theoretically established.

\section 5. Concluding remarks

The results reported in this paper shed some new light
on the nature of non-abelian gauge theories
and the continuum limit in lattice QCD. 
They are obviously incomplete in several respects
and give rise to some interesting questions.
In particular, an all-order analysis of the Wilson flow in
perturbation theory will probably be required in order to achieve a
structural understanding of its renormalization properties.

The one-loop calculation in sect.~2 suggests that
the fields obtained at positive flow time are renormalized
fields for any number of quark flavours.
So far the question was studied numerically
only in the pure gauge theory, 
but it may be encouraging to note that the conjecture is true
in QED, for any charged matter multiplet, 
as a consequence of the gauge Ward identity.

An intriguing aspect of the transformed functional
integral (4.2) is the fact that it is dominated by 
smooth fields. Since the field transformation (4.1) is invertible,
the integral nevertheless encodes all the physics 
described by the theory. 
Some properties
(the division into topological sectors,
for example) however become particularly transparent in this formulation
of the theory,
while its behaviour at high energies is better
discussed in terms of the fundamental gauge field.

\vskip0.3ex
I wish to thank Peter Weisz for a critical reading
of the paper and for checking some of the formulae in sect.~2.
All numerical simulations were performed on a 
dedicated PC cluster at CERN. I am grateful 
to the CERN management for providing the required funds
and to the CERN IT Department for technical support.

\appendix A. Notational conventions

The Lie algebra $\sun$ of $\SUn$ 
may be identified with the linear space of all anti-hermitian
traceless $N\times N$ matrices.
With respect to a 
basis $T^a$, $a=1,\ldots,N^2-1$, of such matrices, 
the elements $X\in\sun$ are given by
$X=X^aT^a$ with real components $X^a$
(repeated group indices are automatically summed over).
The structure constants $f^{abc}$ in
the commutator relation
\equation{
  [T^a,T^b]=f^{abc}T^c
  \enum
}
are real and totally anti-symmetric in the indices if
the normalization condition
\equation{
  \tr\{T^aT^b\}=-\frac{1}{2}\delta^{ab}
  \enum
}
is imposed.
Moreover, $f^{acd}f^{bcd}=N\delta^{ab}$.

Gauge fields in the continuum theory 
take values in the Lie algebra of the gauge group.
Their 
normalization is usually such that
gauge transformations and covariant derivatives 
do not involve the gauge coupling
(see, however, subsect.~2.2).
Lorentz indices $\mu,\nu,\ldots$ in $D$ dimensions 
range from $0$ to $D-1$ and are automatically
summed over when they occur in matching pairs.
The space-time metric is assumed to be Euclidian. In particular,
$p^2=p_{\mu}p_{\mu}$ for any momentum $p$.

The lattice theories considered in this paper are 
set up as usual on a four-dimen\-sional hyper-cubic 
lattice with spacing $a$. In particular, lattice gauge fields $U(x,\mu)$ 
reside on the links
$(x,\mu)$ of the lattice and take values in the gauge group. 
The link differential operators acting on functions
$f(U)$ of the gauge field are
\equation{
  \partial^a_{x,\mu}f(U)=
  {\rmd\over\rmd s}f(\rme^{sX}U)\!\left.{{\vphantom{r\over s}}}\right|_{s=0},
  \hskip1.7em
  X(y,\nu)=\cases{T^a & if $(y,\nu)=(x,\mu)$,\cr
                          \noalign{\vskip1.3ex}
                        0  & otherwise.\cr}
  \enum
}
While these depend on the choice of the generators $T^a$,
the combination
\equation{
  \partial_{x,\mu}f(U)=T^a\partial^a_{x,\mu}f(U)
  \enum
}
can be shown to be basis-independent.

\appendix B. One-loop integrals

The Feynman integrals other than ${\cal E}_0$ which contribute to 
$\langle E\rangle$ at order $g_0^4$ involve an integration over
two momenta, $q$ and $r$, and over at most two 
flow-time parameters. In all cases the integrand is of the 
form
\equation{
  \rme^{-sq^2-ur^2-v(q+r)^2}{P(q,r)\over q^2r^2(q+r)^2}
  \enum
}
where $P(q,r)$ is a Lorentz-invariant polynomial and $s,u,v$ are linear 
combinations of the flow-time parameters.

As already mentioned, these integrals can in principle be evaluated by
substituting the Feynman parameter representation for the propagators
$1/q^2$, $1/r^2$ and $1/(q+r)^2$. 
A more economic computation is, however, always possible 
using the basic integrals
\equation{
  \int_q\rme^{-sq^2}(q^2)^{-\alpha}={s^{\alpha}\over(4\pi s)^{D/2}}
  {\Gamma(D/2-\alpha)\over\Gamma(D/2)},
  \enum
  \nexteq{3.0ex}
  \int_{q,r}{\rme^{-sq^2-ur^2-v(q+r)^2}\over q^2}=
  {2\over(4\pi)^D(D-2)(u+v)}(su+uv+vs)^{1-D/2}.
  \enum
}
The second formula, for example, allows the integral
\equation{
  \int_{q,r}
  {\rme^{-t(q^2+r^2+(q+r)^2)}\over q^2r^2}=
  {1\over(4\pi)^4(2t)^2}\bigl\{8\ln2-4\ln3+\rmO(\eps)\bigr\}
  \enum
}
to be quickly evaluated once the propagator $1/r^2$ is replaced
by its Feynman parameter representation.

A special case are integrals like
\equation{
  \int_0^t\rmd s\int_{q,r}
  {\rme^{-(t+s)(q^2+r^2)-(t-s)(q+r)^2}\over q^2r^2}\,
  qr
  \noenum
  \nexteq{2.0ex}
  {\phantom{\int_0^t\rmd s\int_{q,r}}}
  ={1\over(4\pi)^4(2t)^2}\bigl\{\frac{1}{2}-4\ln2+2\ln3+\rmO(\eps)\bigr\},
  \enum
}
whose integrand is proportional to $qr$. Noting
\equation{
  qr=\frac{1}{2}\left\{(q+r)^2-q^2-r^2\right\},
  \enum
}
the factor can be traded for a differentiation with respect to the
flow-time parameters. Most of the time, this allows one integral over
these parameters to be performed right away and thus leads to 
simpler integrals without factors of $qr$.

\appendix C. Numerical integration of the Wilson flow

On a finite lattice, the space ${\cal G}$ of all gauge fields 
is a finite power of the gauge group and thus itself a Lie group.
The associated Lie algebra $\frak{g}$ coincides with the linear 
space of all link fields
with values in the Lie algebra of the gauge group. From this
abstract point of view,
the flow equation (1.4) is an ordinary first-order differential
equation of the form
\equation{
   \dot{V_t}=Z(V_t)V_t,
   \enum
}
where $V_t\in{\cal G}$ and $Z(V_t)\in\frak{g}$.

The Runge--Kutta scheme 
described in this appendix
obtains the solution of the flow equation at
times $t=n\eps$, $n=1,2,3,\ldots$, recursively,
starting from the initial configuration at $t=0$.
The rule for the integration 
from time $t$ to $t+\eps$ is
\equation{
  W_0=V_t,
  \noenum
  \nexteq{2.5ex}
  W_1=\exp\bigl\{\frac{1}{4}Z_0\bigr\}W_0,
  \noenum
  \nexteq{2.5ex}
  W_2=\exp\bigl\{\frac{8}{9}Z_1-\frac{17}{36}Z_0\bigr\}W_1,
  \noenum
  \nexteq{2.5ex}
  V_{t+\eps}=\exp\bigl\{\frac{3}{4}Z_2
                        -\frac{8}{9}Z_1+\frac{17}{36}Z_0\bigr\}W_2,
  \enum
}
where 
\equation{
  Z_i=\eps Z(W_i),\qquad i=0,1,2.
  \enum
}
Note that this rule is fully explicit.
Moreover, since the gauge field can be overwritten from one
equation to the next, and
since $Z_0$ can be overwritten by
$\frac{8}{9}Z_1-\frac{17}{36}Z_0$, intermediate 
storage space for only one of these latter 
fields is required.

A straightforward calculation
shows that the integration scheme (C.2) 
is accurate up to errors of order $\eps^4$ per step. The total
error of the integration up to a specified flow time thus scales
like $\eps^3$. Empirically one finds that
the integration is numerically stable in the 
direction of positive flow time, 
the integration errors in the link variables being
on the order of $10^{-6}$ if $\eps=0.01$.

\beginbibliography


\bibitem{Wilson}
K. G. Wilson, {\it Confinement of quarks},
Phys. Rev. D10 (1974) 2445


\bibitem{ArnoldI}
V. I. Arnold,
{\it Ordinary differential equations}, 3rd ed. (Springer-Verlag, Berlin, 2008)


\bibitem{Stout}
C. Morningstar, M. Peardon, 
{\it Analytic smearing of SU(3) link variables in lattice QCD},
Phys. Rev. D69 (2004) 054501


\bibitem{TrivMaps}
M. L\"uscher,
{\it Trivializing maps, the Wilson flow and the HMC algorithm},
Commun. Math. Phys. 293 (2010) 899


\bibitem{BardeenEtAl}
W. A. Bardeen, A. J. Buras, D. W. Duke, T. Muta,
{\it Deep inelastic scattering beyond the leading order 
in asymptotically free gauge theories},
Phys. Rev. D18 (1978) 3998


\bibitem{SommerScale}
R. Sommer,
{\it A new way to set the energy scale in lattice gauge theories 
and its applications to the static force and $\alpha_s$ in 
SU(2) Yang--Mills theory},
Nucl. Phys. B411 (1994) 839

\bibitem{GuagnelliEtAl}
M. Guagnelli, R. Sommer, H. Wittig (ALPHA collab.),
{\it Precision computation of a low-energy reference scale in 
quenched lattice QCD},
Nucl. Phys. B535 (1998) 389


\bibitem{DelDebbioTauQ}
L. Del Debbio, H. Panagopoulos, E. Vicari,
{\it $\theta$-dependence of SU(N) gauge theories},
JHEP 0208 (2002) 044


\bibitem{StefanConference}
S. Schaefer, R. Sommer, F. Virotta,
{\it Investigating the critical slowing down of QCD simulations},
PoS (LAT2009) 032


\bibitem{HairerEtAl}
E. Hairer, C. Lubich, G. Wanner,
{\it Geometric numerical integration:
Structure-preserving algorithms for ordinary differential equations},
2nd ed. (Springer, Ber\-lin, 2006)


\bibitem{LambdaParm}
S. Capitani, M. L\"uscher, R. Sommer, H. Wittig
(ALPHA collab.),
{\it Non-pertur\-ba\-tive quark mass renormalization in quenched lattice QCD},
Nucl. Phys. B544 (1999) 669


\bibitem{RitbergenEtAl}
T. van Ritbergen, J. A. M. Vermaseren, S. A. Larin,
{\it The four-loop $\beta$-function in Quantum Chromodynamics},
Phys. Lett. B400 (1997) 379


\bibitem{TopCharge}
M. L\"uscher,
{\it Topology of lattice gauge fields},
Commun. Math. Phys. 85 (1982) 39

\bibitem{PhillipsStone}
A. Phillips, D. Stone, {\it Lattice gauge fields,
principal bundles and the calculation of the topological charge},
Commun. Math. Phys. 103 (1986) 599


\bibitem{IndThm}
P. Hasenfratz, V. Laliena, F. Niedermayer,
{\it The index theorem in QCD with a finite cutoff},
Phys. Lett. B427 (1998) 125

\bibitem{ChiGW}
L. Del Debbio, L. Giusti, C. Pica,
{\it Topological susceptibility in SU(3) gauge theory}
Phys. Rev. Lett. 94 (2005) 032003

\bibitem{ChiInd}
M. L\"uscher,
{\it Topological effects in QCD and the 
problem of short distance singularities},
Phys. Lett. B593 (2004) 296

\bibitem{ChiProj}
L. Giusti, M. L\"uscher,
{\it Chiral symmetry breaking and the Banks--Casher relation 
in lattice QCD with Wilson quarks},
JHEP 0903 (2009) 013


\bibitem{ChiWI}
L. Giusti, G. C. Rossi, M. Testa, G. Veneziano,
{\it The $U_A(1)$ problem on the lattice with Ginsparg--Wilson fermions},
Nucl. Phys. B628 (2002) 234

\bibitem{ChiWII}
L. Giusti, G. C. Rossi, M. Testa,
{\it Topological susceptibility in full QCD with Ginsparg--Wilson fermions},
Phys. Lett. B587 (2004) 157

\endbibliography

\bye